\documentclass[
aps,%
12pt,%
final,%
notitlepage,%
oneside,%
onecolumn,%
nobibnotes,%
nofootinbib,%
superscriptaddress,%
noshowpacs,%
centertags]%
{revtex4}
\allowdisplaybreaks

\begin{document}
\title{The Rarita--Schwinger field: dressing procedure and spin-parity content}
\author{\firstname{A. E. } \surname{Kaloshin}}
\email{kaloshin@physdep.isu.ru}
\author{\firstname{V. P. } \surname{Lomov}}
\email{lomov@physdep.isu.ru} \affiliation{Irkutsk State
University, K.~Marx Str. 1, Irkutsk, 664003, Russia}

\begin{abstract}
We obtain in analytical form the dressed propagator of the massive
Rarita-Schwinger field taking into account all spin components. We found that
the nearest analogy for dressing the Rarita--Schwinger field in
spin--1/2 sector is dressing the two Dirac fermions of opposite parity
with presence of mutual transitions. The calculation of the
self-energy contributions confirms that besides the leading spin--3/2 component
the Rarita--Schwinger field contains also two
spin--1/2  components of different parity.

\end{abstract}


\maketitle

\section{Introduction}

The  description of spin--3/2 particles in (effective) quantum
field theory is usually based on using of a spin-vector field
$\Psi^{\mu}$, called the Rarita--Schwinger field
\cite{Dir,Fie,Rar-Sch}. However, besides the leading spin--3/2
component, this field contains also two spin--1/2 representations
which are usually supposed unphysical. The main difficulties and
paradoxes in its description \cite{JohnSud61,VeZw69,Kob} are in
fact related with existence of "extra" components in $\Psi^{\mu}$
field and with attempts to exclude them in some way. The problem
of consistent description of spin--3/2 particles has a long
history but we mention here only relatively recent works
\cite{Kor,Pas98,Hab,Pas01,Cas,AKS,Pil} where the detailed
discussion may be found. As for applications of this spin-vector
formalism, they are first of all the description of the baryon
resonances production (the most investigated one is
$\Delta(1232)$) in various processes and theoretical description
of gravitino properties.

The most general lagrangian for the free Rarita--Schwinger field
has the following form (see \textit{e.g.}
\cite{Mol,AU,NEK,BDM}):
\begin{flalign}
\mathcal{L}=&\bar{\Psi}
\vphantom{\Psi}^{\mu}\Lambda^{\mu\nu}\Psi^{\nu},\notag\\
\Lambda^{\mu\nu}=&(\hat{p}-M)g^{\mu\nu}+A(\gamma^{\mu}p^{\nu}+
\gamma^{\nu}p^{\mu})+\frac{1}{2}(3A^2+2A+1)\gamma^{\mu}
\hat{p}\gamma^{\nu}+M(3A^2+3A+1)\gamma^{\mu}\gamma^{\nu}.
\label{lagr}
\end{flalign}
Here $M$ is the mass of Rarita--Schwinger field, $A$ is an arbitrary
parameter, $p_{\mu}=i \partial_{\mu}$.

This lagrangian is invariant under the point transformation:
\begin{equation}
\label{symm}
\Psi^{\mu}\to\Psi^{\prime\,\mu}=(g^{\mu\nu}+\alpha\gamma^{\mu}\gamma^{\nu})\Psi^{\nu},
\qquad A\to A^{\prime}=\frac{A-2\alpha}{1+4\alpha},
\end{equation}
with parameter $\alpha\neq -1/4$.

The lagrangian \eqref{lagr} leads to the following equations of motion:
\begin{equation}
\Lambda^{\mu\nu}\Psi^{\nu}=0.
\end{equation}
The free propagator of the Rarita--Schwinger field in a momentum
space obeys the equation:
\begin{equation}
\Lambda^{\mu\nu}G_{0}\mkern -4mu {\vphantom{\Lambda}}^{\nu\rho}=g^{\mu\rho}.
\end{equation}
The expression for the free propagator $G^{\mu\nu}_0$ is well
known (see references above), thus we do not present it here.

As concerned for the dressed propagator, its construction is more
complicated issue and its total expression is unknown up to now.
More exactly, the spin--3/2 contribution may be written out
unambiguously (it has the Breit--Wigner form in case of
$\Delta(1232)$), while the $s=1/2$ contributions present some
problem. Therefore a practical use of the dressed propagator
$G^{\mu\nu}$ needs some approximations in its description. The
standard approximation (\textit{e.g.} \cite{Pas95,KS}) consists in
dressing  the spin--3/2 components only while the rest components
are neglected or considered as bare. Another way to take into
account the spin--1/2 components is a numerical solution of
appearing  system of equations \cite{Kor,AKS}. However, it is
rather difficult to understand in this case, how the unphysical
degrees of freedom are renormalized and whether they are remained
unphysical after the dressing.

The non-leading $s=1/2$ spin contributions from phenomenological
point of view generate the non-resonant background contribution
that interfere with resonance. It is not obvious in advance
whether these contributions are essential for observables. But at
least for $\Delta(1232)$ isobar production in Compton scattering
it was noted \cite{Pas95} that the non-leading contributions  are
necessary for data description.

In this paper%
\footnote{%
The short version of the paper without the spin-parity discussion
was published in \cite{KP} } we derive an analytical expression
for the interacting Rarita--Schwinger field's propagator with
accounting all spin components and discuss its properties. In
particular we identify the spin-parity content of different
contributions  in propagator. The spin--1/2 part has rather
compact form and a crucial point for its deriving is the choosing
of a suitable basis.

\section{Dressed propagator of the Rarita--Schwinger field}

The Dyson--Schwinger equation for the propagator of the
Rarita--Schwinger field has the following form
\begin{equation}
G^{\mu\nu}=G^{\mu\nu}_{0}+G^{\mu\alpha}J^{\alpha\beta}G^{\beta\nu}_{0}.
\end{equation}
Here $G_{0}^{\mu\nu}$ and $G^{\mu\nu}$ are the free and full
propagators respectively, $J^{\mu\nu}$ is a self-energy
contribution. The equation may be rewritten for inverse
propagators as
\begin{equation}
(G^{-1})^{\mu\nu}=(G^{-1}_{0})^{\mu\nu}-J^{\mu\nu}.\label{inverD}
\end{equation}
If to consider the self-energy $J^{\mu\nu}$ as a known value%
\footnote{%
This is the widely used in the resonant physics "rainbow"
approximation, see \textit{e.g.} recent review \cite{MaRo} },
then the problem is reduced to reversing of the relation
\eqref{inverD}. For this procedure it is useful to have some basis
for both propagators and self-energy.
\begin{enumerate}
\item The most natural basis for the spin-tensor $S^{\mu\nu}(p)$
      decomposition is the $\gamma$-matrix one:
      \begin{equation}
      \begin{split}
      S^{\mu\nu}(p)=&g^{\mu\nu}\cdot s_1+p^{\mu}p^{\nu}\cdot s_2+\\
                    &+\hat{p}p^{\mu}p^{\nu}\cdot s_3+\hat{p}g^{\mu\nu}\cdot s_4+p^{\mu}\gamma^{\nu}\cdot s_5+
                       \gamma^{\mu}p^{\nu}\cdot s_6+\\
                    &+\sigma^{\mu\nu}\cdot s_7+\sigma^{\mu\lambda}p^{\lambda}p^{\nu}\cdot s_8+
                       \sigma^{\nu\lambda}p^{\lambda}p^{\mu}\cdot s_9+
                       \gamma^{\lambda}\gamma^{5}\imath\varepsilon^{\lambda\mu\nu\rho}p^{\rho}\cdot s_{10}.
      \end{split}
      \end{equation}
      Here $S^{\mu\nu}$ is an arbitrary spin-tensor depending on the momentum $p$,
      $s_{i}(p^2)$ are the Lorentz invariant coefficients, and
      $\sigma^{\mu\nu}=\frac{1}{2}[\gamma^{\mu},\gamma^{\nu}]$.
      Altogether there are ten independent components in the decomposition
      of $S^{\mu\nu}(p)$, if parity is conserved.

      It is known that the $\gamma$-matrix decomposition is complete,
      the coefficients $s_{i}$ are free of kinematic
      singularities and constraints, and their calculation is rather
      simple. However this basis is inconvenient at multiplication and
      reversing of the spin-tensor $S^{\mu\nu}(p)$ because the basis
      elements are not orthogonal to each other. As a result the
      reversing of the spin-tensor $S^{\mu\nu}(p)$ leads to a system of 10
      equations for the coefficients.

\item There is another basis used  in  consideration
      of the dressed propagator \cite{Kor,AKS,Pas95} $G^{\mu\nu}$. It is
      constructed from the following set of operators%
      \footnote{%
      We changed here, for convenience, the normalization of $\mathcal{P}^{1/2}_{21}$,
      $\mathcal{P}^{1/2}_{12}$.
      }
      \cite{BDM,Pas95,PvanNie}
      \begin{flalign}
      (\mathcal{P}^{3/2})^{\mu\nu}=&g^{\mu\nu}-\frac{2}{3}\frac{p^{\mu}p^{\nu}}{p^2}-
                                     \frac{1}{3}\gamma^{\mu}\gamma^{\nu}+\frac{1}{3p^2}(\gamma^{\mu}p^{\nu}-
                                     \gamma^{\nu}p^{\mu})\hat{p},\notag\\
      (\mathcal{P}^{1/2}_{11})^{\mu\nu}=&\frac{1}{3}\gamma^{\mu}\gamma^{\nu}-
                                          \frac{1}{3}\frac{p^{\mu}p^{\nu}}{p^2}-\frac{1}{3p^2}(\gamma^{\mu}p^{\nu}-
                                          \gamma^{\nu}p^{\mu})\hat{p},\notag\\
      (\mathcal{P}^{1/2}_{22})^{\mu\nu}=&\frac{p^{\mu}p^{\nu}}{p^2},\notag\\
      (\mathcal{P}^{1/2}_{21})^{\mu\nu}=&\sqrt{\frac{3}{p^2}}\cdot\frac{1}{3p^2}(-p^{\mu}+\gamma^{\mu}\hat{p})
                                          \hat{p}p^{\nu},\notag\\
      (\mathcal{P}^{1/2}_{12})^{\mu\nu}=&\sqrt{\frac{3}{p^2}}\cdot\frac{1}{3p^2}p^{\mu}(-p^{\nu}+\gamma^{\nu}\hat{p})
                                          \hat{p}.
      \label{present}
      \end{flalign}
      The first three operators
      $\mathcal{P}^{3/2}$, $\mathcal{P}^{1/2}_{11}$, $\mathcal{P}^{1/2}_{22}$
      are the projection operators while
      $\mathcal{P}^{1/2}_{21}$, $\mathcal{P}^{1/2}_{12}$ are nilpotent ones.
      As for their physical meaning, it is obvious that
      $\mathcal{P}^{3/2}$ corresponds to spin--3/2. The remaining
      operators should describe two spin--1/2 representations and
      transitions between them.
      Let us rewrite the operators \eqref{present} to make their properties more obvious:
      \begin{flalign}
      (\mathcal{P}^{3/2})^{\mu\nu}=&g^{\mu\nu}-(\mathcal{P}^{1/2}_{11})^{\mu\nu}-(\mathcal{P}^{1/2}_{22})^{\mu\nu},
                                       \notag\\
      (\mathcal{P}^{1/2}_{11})^{\mu\nu}=&3\pi^{\mu}\pi^{\nu},\notag\\
      (\mathcal{P}^{1/2}_{22})^{\mu\nu}=&\frac{p^{\mu}p^{\nu}}{p^2},\notag\\
      (\mathcal{P}^{1/2}_{21})^{\mu\nu}=&\sqrt{\frac{3}{p^2}}\cdot\pi^{\mu}p^{\nu},\notag\\
      (\mathcal{P}^{1/2}_{12})^{\mu\nu}=&\sqrt{\frac{3}{p^2}}\cdot
      p^{\mu}\pi^{\nu}.
      \end{flalign}
      Here we introduced the vector
      \begin{equation}
      \pi^{\mu}=\frac{1}{3p^2}(-p^{\mu}+\gamma^{\mu}\hat{p})\hat{p}
      \end{equation}
      with the following properties:
      \begin{equation}
      (\pi p)=0,\quad (\gamma\pi)=(\pi\gamma)=1,\quad (\pi\pi)=\frac{1}{3},\quad \hat{p}\pi^{\mu}=-\pi^{\mu}\hat{p}.
      \end{equation}
      The set of operators \eqref{present} can be used to decompose the
      considered spin-tensor as follows \cite{Kor,AKS}:
      \begin{equation}
      \label{p-expan}
      \begin{split}
      S^{\mu\nu}(p)&=(S_{1}+S_{2}\hat{p})(\mathcal{P}^{3/2})^{\mu\nu}+
      (S_{3}+S_{4}\hat{p})(\mathcal{P}^{1/2}_{11})^{\mu\nu}+
      (S_{5}+S_{6}\hat{p})(\mathcal{P}^{1/2}_{22})^{\mu\nu}+ \\
      &+(S_{7}+S_{8}\hat{p})(\mathcal{P}^{1/2}_{21})^{\mu\nu}+
      (S_{9}+S_{10}\hat{p})(\mathcal{P}^{1/2}_{12})^{\mu\nu}.
      \end{split}
      \end{equation}
      Let us call this basis as $\hat{p}$-basis. It is more convenient
      at multiplication since the spin--3/2 components
      $\mathcal{P}^{3/2}$ have been separated from spin--1/2 ones.
      However, the spin--1/2 components as before are not orthogonal
      between themselves and we come to a system of 8 equations when
      inverting the \eqref{inverD}. Another feature of decomposition
      \eqref{p-expan} is existence of the poles $1/p^2$ in different
      terms. So, to avoid this unphysical singularity, we should impose
      some constraints on the coefficients at zero point.
\item Let us construct the basis which is the most convenient
      at multiplication of spin-tensors. This basis is built from the
      operators \eqref{present} and the projection operators
      $\Lambda^{\pm}$
      \begin{equation}
      \Lambda^{\pm}=\frac{\sqrt{p^2}\pm\hat{p}}{2\sqrt{p^2}},
      \label{project}
      \end{equation}
      where we assume $\sqrt{p^2}$ to be the first branch of analytical function.
      Ten elements of this basis look as
      \begin{flalign}
      \mathcal{P}_{1}=&\Lambda^{+}\mathcal{P}^{3/2},\,&
      \mathcal{P}_{3}=&\Lambda^{+}\mathcal{P}^{1/2}_{11},\,&
      \mathcal{P}_{5}=&\Lambda^{+}\mathcal{P}^{1/2}_{22},\,&
      \mathcal{P}_{7}=&\Lambda^{+}\mathcal{P}^{1/2}_{21},\,&
      \mathcal{P}_{9}=&\Lambda^{+}\mathcal{P}^{1/2}_{12},\notag\\
      \mathcal{P}_{2}=&\Lambda^{-}\mathcal{P}^{3/2},\,&
      \mathcal{P}_{4}=&\Lambda^{-}\mathcal{P}^{1/2}_{11},\,&
      \mathcal{P}_{6}=&\Lambda^{-}\mathcal{P}^{1/2}_{22},\,&
      \mathcal{P}_{8}=&\Lambda^{-}\mathcal{P}^{1/2}_{21},\,&
      \mathcal{P}_{10}=&\Lambda^{-}\mathcal{P}^{1/2}_{12},
      \label{L-basis}
      \end{flalign}
      where tensor indices are omitted. We will call \eqref{L-basis} as
      the $\Lambda$-basis.

      The decomposition of a spin-tensor in this basis has the following form:
      \begin{equation}
      S^{\mu\nu}(p)=\sum_{i=1}^{10}\mathcal{P}^{\mu\nu}_{i}\bar{S}_{i}(p^2).
      \label{l-expan}
      \end{equation}

      The coefficients $\bar{S}_{i}$ are calculated in analogy with $\gamma$-matrix
      decomposition. Besides, we found (with computer analytical calculations) the matrix
      relating the $\Lambda$-basis with the $\gamma$-matrix basis and convinced
      ourselves that this matrix is not singular. Therefore the elements of this basis \eqref{L-basis}
      are independent and the basis is complete.
      It is easy to connect the expansion coefficients \eqref{p-expan} and \eqref{l-expan}
      between themselves.
      \begin{align}
      \bar{S}_{1}&=S_1+\sqrt{p^2}S_2, &
      \bar{S}_{2}&=S_1-\sqrt{p^2}S_2,\notag\\
      \bar{S}_{3}&=S_3+\sqrt{p^2}S_4, &
      \bar{S}_{4}&=S_3-\sqrt{p^2}S_4,\notag\\
      \bar{S}_{5}&=S_5+\sqrt{p^2}S_6, &
      \bar{S}_{6}&=S_5-\sqrt{p^2}S_6,\\
      \bar{S}_{7}&=S_7+\sqrt{p^2}S_8, &
      \bar{S}_{8}&=S_7-\sqrt{p^2}S_8, \notag\\
      \bar{S}_{9}&=S_9+\sqrt{p^2}S_{10}, &
      \bar{S}_{10}&=S_9-\sqrt{p^2}S_{10},\notag
      \end{align}

      The $\Lambda$-basis has very simple multiplicative properties
      which are represented in the Table~\ref{tt}.

      \begin{table}[ht]

      \begin{center}
       \begin{tabular}{c|cccccccccc}
      \qquad           & $\mathcal{P}_1$ & $\mathcal{P}_2$ & $\mathcal{P}_3$ & $\mathcal{P}_4$ & $\mathcal{P}_5$ & $\mathcal{P}_6$ & $\mathcal{P}_7$ & $\mathcal{P}_8$ & $\mathcal{P}_9$ & $\mathcal{P}_{10}$\\
      \hline
      $\mathcal{P}_1$    & $\mathcal{P}_1$ & 0&0&0&0&0&0&0&0&0\\
      $\mathcal{P}_2$    &0&$\mathcal{P}_2$&0&0&0&0&0&0&0&0\\
      $\mathcal{P}_3$    &0&0&$\mathcal{P}_3$&0&0&0&$\mathcal{P}_7$&0&0&0\\
      $\mathcal{P}_4$    &0&0&0&$\mathcal{P}_4$&0&0&0&$\mathcal{P}_8$&0&0\\
      $\mathcal{P}_5$    &0&0&0&0&$\mathcal{P}_5$&0&0&0&$\mathcal{P}_9$&0\\
      $\mathcal{P}_6$    &0&0&0&0&0&$\mathcal{P}_6$&0&0&0&$\mathcal{P}_{10}$\\
      $\mathcal{P}_7$    &0&0&0&0&0&$\mathcal{P}_7$&0&0&0&$\mathcal{P}_3$\\
      $\mathcal{P}_8$    &0&0&0&0&$\mathcal{P}_8$&0&0&0&$\mathcal{P}_4$&0\\
      $\mathcal{P}_9$    &0&0&0&$\mathcal{P}_9$&0&0&0&$\mathcal{P}_5$&0&0\\
      $\mathcal{P}_{10}$ &0&0&$\mathcal{P}_{10}$&0&0&0&$\mathcal{P}_6$&0&0&0\\
      \end{tabular}
      \end{center}
      \caption{Properties of the $\Lambda$-basis at multiplication.}
      \label{tt}
      \end{table}

      The first six basis elements are projection operators, while the
      remaining four elements are nilpotent. We are convinced by direct
      calculations that there are no other projection operators besides
      indicated.
\end{enumerate}

Now we can return to the Dyson--Schwinger equation \eqref{inverD}.
Let us denote the inverse dressed propagator $(G^{-1})^{\mu\nu}$
and free one $(G^{-1}_{0})^{\mu\nu}$ by $S^{\mu\nu}$ and $S^{\mu\nu}_{0}$
respectively.
Decomposing the $S^{\mu\nu}$, $S_{0}^{\mu\nu}$ and $J^{\mu\nu}$ in $\Lambda$-basis
according to \eqref{l-expan} we reduce the equation \eqref{inverD} to set of
equations for the scalar coefficients
\begin{equation*}
\bar{S}_{i}(p^2)=\bar{S}_{0i}(p^2)+\bar{J}_{i}(p^2)
\end{equation*}
The values $\bar{S}_{i}$ are defined by the bare propagator and the
self-energy and may be considered as known.

The dressed propagator also can be found in such form
\begin{equation}
G^{\mu\nu}=\sum_{i=1}^{10}\mathcal{P}^{\mu\nu}_{i}\cdot\bar{G}_{i}(p^2)
\label{ddecomp}
\end{equation}
The existing $6$ projection operators take part in the decomposition of $g^{\mu\nu}$:
\begin{equation}
g^{\mu\nu}=\sum_{i=1}^{6}\mathcal{P}^{\mu\nu}_{i}.
\end{equation}
Now solving the equation
\begin{equation*}
G^{\mu\nu}S^{\nu\lambda}=g^{\mu\lambda}
\end{equation*}
in $\Lambda$-basis, we obtain a set of equations for the scalar coefficients
$\bar{G}_{i}$.
\begin{gather}
\begin{array}{ccc}
\bar{G}_1 \bar{S}_1 &=& 1, \\
\bar{G}_2 \bar{S}_2 &=& 1,
\end{array}
\notag\\
\begin{array}{ccc}
\bar{G}_3 \bar{S}_3 + \bar{G}_7 \bar{S}_{10}      &=& 1,       \\
\bar{G}_{3} \bar{S}_{7} + \bar{G}_{7} \bar{S}_{6} &=& 0,
\end{array}
\qquad
\begin{array}{ccc}
\bar{G}_4 \bar{S}_4 + \bar{G}_8 \bar{S}_{9}       &=& 1,        \\
\bar{G}_{4} \bar{S}_{8} + \bar{G}_{8} \bar{S}_{5} &=& 0,
\end{array}
\\
\begin{array}{ccc}
\bar{G}_5 \bar{S}_5 + \bar{G}_9 \bar{S}_{8}       &=& 1,        \\
\bar{G}_{5} \bar{S}_{9} + \bar{G}_{9} \bar{S}_{4} &=& 0,
\end{array}
\qquad
\begin{array}{ccc}
\bar{G}_6 \bar{S}_6 + \bar{G}_{10} \bar{S}_{7}      &=& 1,     \\
\bar{G}_{6} \bar{S}_{10} + \bar{G}_{10} \bar{S}_{3} &=& 0 .
\end{array}
\notag
\end{gather}

The equations are easy to solve:
\begin{flalign}
\label{solve}
\bar{G}_1&=\frac{1}{\bar{S}_1},\quad \bar{G}_2=\frac{1}{\bar{S}_2},\notag\\
\bar{G}_3&=\frac{\bar{S}_6}{\Delta_1},\quad
\bar{G}_4=\frac{\bar{S}_5}{\Delta_2},\quad
\bar{G}_5=\frac{\bar{S}_4}{\Delta_2},\quad
\bar{G}_6=\frac{\bar{S}_3}{\Delta_1},\\
\bar{G}_7&=\frac{-\bar{S}_7}{\Delta_1},\quad
\bar{G}_8=\frac{-\bar{S}_8}{\Delta_2},\quad
\bar{G}_9=\frac{-\bar{S}_9}{\Delta_2},\quad
\bar{G}_{10}=\frac{-\bar{S}_{10}}{\Delta_1},\notag
\end{flalign}
where
\begin{equation}
\Delta_1=\bar{S}_{3}\bar{S}_{6}-\bar{S}_{7}\bar{S}_{10},\qquad
\Delta_2=\bar{S}_{4}\bar{S}_{5}-\bar{S}_{8}\bar{S}_{9}.
\end{equation}
The $\bar{G}_{1}$, $\bar{G}_{2}$ terms which describe
the spin--3/2 have the usual resonant form and could be obtained
from \eqref{p-expan} as well. As for $\bar{G}_{3} -
\bar{G}_{10}$ coefficients which describe the spin--1/2
contributions, they have a more complicated structure. Let us
consider the denominators of \eqref{solve} in more details.
\begin{flalign}
\Delta_1=&\bar{S}_{3}\bar{S}_{6}-\bar{S}_{7}\bar{S}_{10}=(S_3+\sqrt{p^2}S_4)(S_5-\sqrt{p^2}S_6)-
         (S_7+\sqrt{p^2}S_8)(S_9-\sqrt{p^2}S_{10}),\notag\\
\Delta_2=&\Delta_1(\sqrt{p^2}\to-\sqrt{p^2}).\label{denoms}
\end{flalign}
The appearance of $\sqrt{p^2}$ factor is typical for fermions --- see below and
this apparent branch point $\sqrt{p^2}$ is canceled in total expression for the dressed
Rarita--Schwinger propagator \eqref{solve}.

Thus we obtained the simple analytical expression \eqref{solve}
for the interacting Rarita-Schwinger field propagator which
accounts  all spin components.
 The new moment here is a closed expression
for spin--1/2 sector where we can expect the dressing of two
spin--1/2 components with mutual transitions. To derive it we
introduced the spin-tensor basis \eqref{L-basis} with very simple
multiplicative properties. This basis is singular and it seems
unavoidable (recall the vector field case). Nevertheless the
singularity of a basis is not so big obstacle in its use though it
needs additional constrains on coefficients. We did not suppose
here any symmetry properties of the self-energy $J^{\mu\nu}$
restricting ourselves by general case. Of course the concrete form
of interaction will lead to some symmetry of $J^{\mu\nu}$ and it
will be important at renormalization.

\section{Dressing of Dirac fermions}

The obtained answer for the interacting Rarita-Schwinger field's propagator
has rather unusual structure, so before
renormalization it would be useful to clarify the physical meaning
of the formulae. In search for nearest analogy for dressing of
Rarita--Schwinger field  we consider below few examples for
dressing of Dirac fermions. We will use the projection operators
$\Lambda^{\pm}$ \eqref{project} which has been appeared in
consideration of Rarita--Schwinger field. We found them very
convenient in case of Dirac fermions also.

\subsection{Dressed fermion propagator}

The dressed fermion propagator $G(p)$ is the solution of the
Dyson--Schwinger equation
\begin{equation}
\label{one-f-DS}
G(p)=G_{0}+G\Sigma G_{0},
\end{equation}
where $G_{0}$ is a bare propagator and $\Sigma$ is a self-energy contribution.

Let us introduce here new notations
to emphasize the analogy with Rarita--Schwinger field.
\begin{equation}
\mathcal{P}_{1}=\Lambda^{+}=\frac{\sqrt{p^2}+\hat{p}}{2\sqrt{p^2}},\qquad
\mathcal{P}_{2}=\Lambda^{-}=\frac{\sqrt{p^2}-\hat{p}}{2\sqrt{p^2}}.
\end{equation}
Decomposition of any $4\times4$ matrix depending on one momentum
$p$ has the form
\begin{equation}
\label{p2-expan} S(p)=\sum_{M=1}^{2}\mathcal{P}_{M}\bar{S}^{M}.
\end{equation}
The Dyson--Schwinger equation \eqref{one-f-DS} in this basis takes
form
\begin{equation}
\bar{G}^{M}=\bar{G}^{M}_{0}+
\bar{G}^{M} \bar{\Sigma}^{M} \bar{G}^{M}_{0},   \qquad  M=1,2 ,
\end{equation}
or, equivalently,
\begin{equation}
(\bar{G}^{M})^{-1} = (\bar{G}^{M}_0)^{-1} - \bar{\Sigma}^M   .
\end{equation}
More detailed answer is
\begin{flalign}
\big(\bar{G}^{M=1}\big)^{-1} = \big(\bar{G}^{M=1}_0\big)^{-1} - \bar{\Sigma}^{M=1}=
-m_{0}-A(p^2)+\sqrt{p^2}\big(1-B(p^2)\big),\\
\big(\bar{G}^{M=2}\big)^{-1} = \big(\bar{G}^{M=2}_0\big)^{-1} - \bar{\Sigma}^{M=2}=
-m_{0}-A(p^2)-\sqrt{p^2}\big(1-B(p^2)\big),
\end{flalign}
where $A$,$B$ are usually used components of the self-energy
contribution
\begin{equation*}
\begin{split}
\Sigma(p)=A(p^2)+\hat{p}B(p^2)=\Lambda^{+}\Sigma^{1}+\Lambda^{-}\Sigma^{2},\\
\Sigma^{1}=A+\sqrt{p^2}B,\qquad  \Sigma^{2}=A-\sqrt{p^2}B.
\end{split}
\end{equation*}

The renormalization has some  features because of using of
$\Lambda^{\pm}$ operators, however the final answer coincides with
standard (see Appendix).

Let us look at the self-energy contribution $\Sigma(p)$. As an
example we shall consider the dressing of baryon resonance
$N^{\prime}$ $(J^{P}=1/2^{\pm})$  due to interaction with $\pi N$
system. Interaction lagrangian is of the form
\footnote{%
The isotopic is not essential for our purpose, so we do not write
these indices.}
\begin{equation}
{L}_{int}=g\bar{\Psi}^{\ \prime}(x)\gamma^5\Psi(x)\cdot\phi(x)+h.c.
\quad
\text{for}\quad
N^{\prime}=1/2^{+}
\end{equation}
and
\begin{equation}
{L}_{int}=g\bar{\Psi}^{\ \prime}(x)\Psi(x)\cdot\phi(x)+h.c.
\quad
\text{for}\quad
N^{\prime}=1/2^{-} .
\end{equation}\\[-1cm]

Let us write down the one-loop self-energy contribution.

\medskip
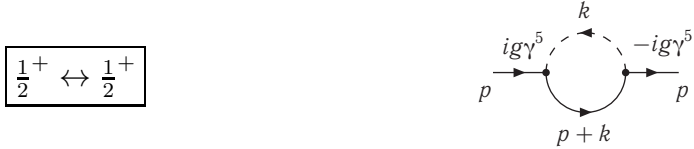
\raisebox{\rbcorh}{\fbox{$\frac{1}{2}^{+}\leftrightarrow\frac{1}{2}^{+}$}}
\hspace*{30mm}
\includegraphics*{\imagepath{loop11}}

\begin{equation}
\label{one-f-loop}
\Sigma(p)=-ig^2 \int \frac{d^4k}{(2\pi)^4}\ \gamma^5\frac{1}{\hat{p}+\hat{k}-m_N}\gamma^5
\frac{1}{k^2-\mpi^2}=I\cdot A(p^2)+\hat{p}B(p^2)
\end{equation}
The loop discontinuity is calculated according to Landau--Cutkosky rule%
\footnote{%
This is a way to avoid unphysical singularities:  to renormalize
the $A$, $B$ components and then to calculate $\Sigma^{\pm}$. }
\begin{equation}
\label{one-f-lp}
\Delta A=\frac{ig^2m_N}{(2\pi)^2}I_{0},
\qquad
\Delta B=-\frac{ig^2}{(2\pi)^2}I_{0}\frac{p^2+m_N^2-\mpi^2}{2p^2}.
\end{equation}
Here $I_{0}$ is the basic integral without indices
\begin{equation}
\label{ibase}
\begin{split}
I_{0}&=\int d^4k\delta\big(k^2-\mpi^2\big)\delta\big((p+k)^2-m_N^2\big)=\theta\big(p^2-(m_N+\mpi)^2\big)
\frac{\pi}{2}\sqrt{\frac{\lambda\big(p^2,m_N^2,\mpi^2\big)}{\big(p^2\big)^2}},
\end{split}
\end{equation}
where $\lambda\big(a,b,c\big)=\big(a-b-c\big)^2-4bc$.

From the parity conservation one can see that in the transition
$N^{\prime}(1/2^+)\to N(1/2^+)+\pi(0^{-})$ the $\pi N$ pair has
the orbital momentum $l=1$. But according to threshold
quantum-mechanical theorems \cite{Baz}, the imaginary part of a
loop should behave as $q^{2l+1}$ at $q\to0$, which does not
correspond to \eqref{one-f-lp}.
\footnote{%
$q$ is a spatial momentum of $\pi N$ pair in CMS
\begin{equation*}
q^2=\frac{\lambda\big(p^2,m_N^2,\mpi^2\big)}{4p^2}=\frac{\big[p^2-(m_N+\mpi)^2\big]\big[p^2-(m_N-\mpi)^2\big]}
{4p^2}
\end{equation*}
} Let us calculate the imaginary part of $\Sigma^{M}$ component
according to \eqref{one-f-lp}
\begin{equation}
\label{im-ofloop}
\begin{split}
\IM\Sigma^{1}&= \IM\big(A+\sqrt{p^2}B\big)=\frac{g^2I_{0}}{4\sqrt{p^2}(2\pi)^2}
\Big(\sqrt{p^2}-m_N-\mpi\Big)\Big(\sqrt{p^2}-m_N+\mpi\Big)\sim q^3,\\
\IM\Sigma^{2}&= \IM\big(A-\sqrt{p^2}B\big)=-\frac{g^2I_{0}}{4\sqrt{p^2}(2\pi)^2}
\Big[\big(\sqrt{p^2}+m_N\big)^2-\mpi^2\Big]\sim q^1.
\end{split}
\end{equation}
One can see that the $\Sigma^{1}$ component (accompanied by the
$\Lambda^{+}$ projector) demonstrates the proper threshold
behavior.

\medskip
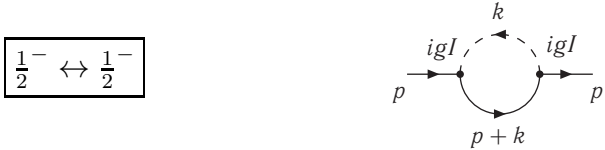
\raisebox{\rbcorh}{\fbox{$\frac{1}{2}^{-}\leftrightarrow\frac{1}{2}^{-}$}}
\hspace*{30mm}
\includegraphics*{\imagepath{loop22}}

\begin{equation}
\label{rec-of-loop}
\begin{split}
\Sigma(p)&=ig^2 \int
\frac{d^4k}{(2\pi)^4}\ \frac{1}{\hat{k}+\hat{p}-m_N}\cdot\frac{1}{k^2-\mpi^2}=IA(p^2)+
\hat{p}B(p^2),\\
\Delta A&=-i\frac{g^2m_N}{(2\pi)^2}I_{0},\qquad
\Delta B=\frac{-ig^2}{(2\pi)^2}I_{0}\frac{p^2+m_N^2-\mpi^2}{2p^2}
\end{split}
\end{equation}
Again the  $\Sigma^{1}$ component demonstrates the proper orbital
momentum ($l=0$ in this case)
\begin{equation}
\label{im-ofloop-2}
\begin{split}
\IM\Sigma^{1}&=-\frac{g^2I_{0}}{4\sqrt{p^2}(2\pi)^2}
\Big[\big(\sqrt{p^2}+m_N\big)^2-\mpi^2\Big] \sim q^1,\\
\IM\Sigma^{2}&=\frac{g^2I_{0}}{4\sqrt{p^2}(2\pi)^2}\big(\sqrt{p^2}-m_N-\mpi\big)
\big(\sqrt{p^2}-m_N+\mpi\big) \sim q^3
\end{split}
\end{equation}

So the considered examples indicate that the correct threshold
behavior (\textit{i.e.} correct parity) demonstrates only
$\Sigma^{1}$ component which has the
$1\big/\big(\sqrt{p^2}-m\big)$ pole. As for another component
$\Sigma^{2}$, which has pole of the form
$1\big/\big(-\sqrt{p^2}-m\big)$, it has the opposite parity. The
appearance of opposite parity contributions in a propagator is
quite natural since the fermion and anti-fermion parities are
opposite.

\subsection{Fermion dressing with parity violation}

Let us consider a dressing of  the fermion state with parity
violation. Such situation, arises, in particular for dressing of
the $t$-quark propagator. Dyson--Schwinger equation has the same
form but the self-energy contribution $\Sigma$ contains the parity
violating terms
\begin{equation}
\Sigma(p)=A(p^2)+\hat{p}B(p^2)+\gamma^5C(p^2)+\hat{p}\gamma^5D(p^2).
\end{equation}
Now the basis will contain four operators:
\begin{equation}
\label{p4-bas}
\mathcal{P}_{1}=\Lambda^{+},\quad
\mathcal{P}_{2}=\Lambda^{-},\quad
\mathcal{P}_{3}=\Lambda^{+}\gamma^5,\quad
\mathcal{P}_{4}=\Lambda^{-}\gamma^5,
\end{equation}
where $\mathcal{P}_{1,2}$ are projection operators  and
$\mathcal{P}_{3,4}$ are nilpotent ones. The expansion of any
$\gamma$-matrix depending on $p$ now has the form (compare with
\eqref{p2-expan})
\begin{equation}
\label{p4-expan}
S(p)=\sum_{M=1}^{4}\mathcal{P}^{M}\bar{S}^{M}.
\end{equation}

This set of operators has simple multiplication properties (see
Table~\ref{mult1}).
\begin{table}[htb]
\begin{center}
\begin{tabular}{c|cccc}
{}&$\mathcal{P}_{1}$&$\mathcal{P}_{2}$&$\mathcal{P}_{3}$&$\mathcal{P}_{4}$\\ \hline
$\mathcal{P}_{1}$&$\mathcal{P}_{1}$&0&$\mathcal{P}_{3}$&0\\
$\mathcal{P}_{2}$&0&$\mathcal{P}_{2}$&0&$\mathcal{P}_{4}$\\
$\mathcal{P}_{3}$&0&$\mathcal{P}_{3}$&0&$\mathcal{P}_{1}$\\
$\mathcal{P}_{4}$&$\mathcal{P}_{4}$&0&$\mathcal{P}_{2}$&0\\
\end{tabular}
\end{center}
\caption{Multiplication properties of the basis \eqref{p4-bas} operators}
\label{mult1}
\end{table}

Let us denote the inverse dressed and bare propagators as $S(p)$
and $S_{0}(p)$ respectively. With using of the basis
\eqref{p4-expan} the Dyson--Schwinger equation is reduced to
\begin{equation*}
\bar{S}^{M}=\big(\bar{S}_{0}\big)^{M}-\bar{\Sigma}^{M},\qquad
M=1,\dots,4 .
\end{equation*}
So the problem is reduced to reversing of the known $S(p)$ matrix
\begin{equation}
\label{label1}
\Big(\sum_{M=1}^{4}\mathcal{P}_{M}\bar{G}^{M}\Big)
\Big(\sum_{L=1}^{4}\mathcal{P}_{L}\bar{S}^{L}\Big)=
\mathcal{P}_{1}+\mathcal{P}_{2}.
\end{equation}
We obtain the set of equations on the unknown coefficients $\bar{G}^{M}$
\begin{equation}
\label{tf-DS-split}
\begin{split}
\begin{array}{rcl}
\bar{G}^{1}\bar{S}^{1}+\bar{G}^{3}\bar{S}^{4} &=& 1\\
\bar{G}^{2}\bar{S}^{2}+\bar{G}^{4}\bar{S}^{3} &=& 1
\end{array}
\qquad\qquad
\begin{array}{rcl}
\bar{G}^{1}\bar{S}^{3}+\bar{G}^{3}\bar{S}^{2} &=& 0 \\
\bar{G}^{4}\bar{S}^{1}+\bar{G}^{2}\bar{S}^{4} &=& 0,
\end{array}
\end{split}
\end{equation}
which are easy to solve. The answer is
\begin{equation}
\label{tf-DS-split-sol}
\bar{G}_{1}=\frac{\bar{S}_{2}}{\Delta},\quad
\bar{G}_{2}=\frac{\bar{S}_{1}}{\Delta},\quad
\bar{G}_{3}=-\frac{\bar{S}_{3}}{\Delta},\quad
\bar{G}_{4}=-\frac{\bar{S}_{4}}{\Delta},
\end{equation}
where $\Delta=\bar{S}_{1}\bar{S}_{2}-\bar{S}_{3}\bar{S}_{4}$.

This example resembles the dressing of the Rarita--Schwinger field
by its algebraic structure (compare Tables 1,2)  but it contains
only few degrees of freedom.

\subsection{Joint dressing of two fermions of the same parities}

Let we have two bare fermion states $N^{\prime}$,
$N^{\prime\prime}$ which are  dressed in presence of mutual
transitions. Suppose that we have two fermions of the same parity
and there is no parity violation in lagrangian. The
Dyson--Schwinger equation acquires matrix indices
\begin{equation}
G_{ij}=\big(G_{0}\big)_{ij}+G_{ik}\Sigma_{kl}\big(G_{0}\big)_{lj},
\qquad
i,j,k,l=1,2.
\label{two-f-DS}
\end{equation}
Each element of this equation has also the $\gamma$-matrix indices
which are not shown.

Using the expansion \eqref{p2-expan} (there is no parity violation
so the basis contains only two elements) we reduce equation
\eqref{two-f-DS} to independent equations on $\bar{G}^{M}$
\begin{equation}
\big(\bar{G}^{M}\big)_{ij}=\big(\bar{G}^{M}_{0}\big)_{ij}+
\big(\bar{G}^{M}\big)_{ik}\big(\bar{\Sigma}^{M}\big)_{kl}\big(\bar{G}^{M}_{0}\big)_{lj},   \qquad  M=1,2.
\label{tfDS-comp}
\end{equation}
Let us rewrite \eqref{tfDS-comp} in  the matrix form
\begin{equation}
\begin{split}
\bar{G}^{M}&=\bar{G}^{M}_{0}+\bar{G}^{M}
\bar{\Sigma}^{M}\bar{G}^{M}_{0},
\quad M=1,2,
\end{split}
\end{equation}
and write down its solution
\begin{equation}
\label{tf-one-par}
\begin{split}
G^{M}&=\Big[\big(\bar{G}^{M}_{0}\big)^{-1}-\bar{\Sigma}^{M}\Big]^{-1}=\left(
\begin{array}{cc}
\big(\bar{G}^{M}_{0}\big)^{-1}_{11}-\bar{\Sigma}^{M}_{11} & -
\bar{\Sigma}^{M}_{12}\\
-\bar{\Sigma}^{M}_{21} & \big(\bar{G}^{M}_{0}\big)^{-1}_{22}-
\bar{\Sigma}^{M}_{22}
\end{array}\right)^{-1}=\\[3mm]
&=\frac{1}{\Delta^{M}}\left(
\begin{array}{cc}
\big(\bar{G}^{M}_{0}\big)^{-1}_{22}-\bar{\Sigma}^{M}_{22} & -\bar{\Sigma}^{M}_{12}\\
-\bar{\Sigma}^{M}_{21} & \big(\bar{G}^{M}_{0}\big)^{-1}_{11}-\bar{\Sigma}^{M}_{11}
\end{array}\right),\\[3mm]
\Delta^{M}&=\Big[\big(\bar{G}_{0}^{M}\big)^{-1}_{11}-\bar{\Sigma}^{M}_{11}\Big]
\Big[\big(\bar{G}^{M}_{0}\big)^{-1}_{22}-\bar{\Sigma}^{M}_{22}\Big]-
\bar{\Sigma}^{M}_{12}\bar{\Sigma}^{M}_{21}.
\end{split}
\end{equation}

Now we will calculate the loop contributions $\Sigma_{ij}$ for
above considered example of baryon dressing  by $\pi N$
intermediate state. For case of the dressing of two states
$N^{\prime}$, $N^{\prime\prime}$ of the same parity the
self-energy contribution coincides with above written out
\eqref{one-f-loop}, \eqref{im-ofloop} besides the coupling
constants.
\begin{equation}
\label{two-f-loop-neg}
\begin{split}
\Sigma_{ij}= ig_{i}g_{j} \int
\frac{d^4k}{(2\pi)^4}\ \gamma^5\frac{1}{\hat{p}+\hat{k}-m}\gamma^5
\frac{1}{k^2-\mpi^2} \qquad \text{for}\quad
N^{\prime}, N^{\prime\prime}=1/2^{+} , \\
\Sigma_{ij}=ig_{i}g_{j} \int
\frac{d^4k}{(2\pi)^4}\ \frac{1}{\hat{p}+\hat{k}-m}\frac{1}{k^2-\mpi^2}
\qquad \text{for}\quad
N^{\prime}, N^{\prime\prime}=1/2^{-} .
\end{split}
\end{equation}

So the above conclusions about the threshold behavior of imaginary parts
\eqref{im-ofloop}, \eqref{im-ofloop-2} are kept in this case also.

\subsection{Joint dressing of two fermions of opposite parities}

Let us consider the nearest analogy to the Rarita--Schwinger filed: the joint dressing of
two fermions of the different parity $1/2^{\pm}$.
We shall suppose that interaction conserves the parity.

As in previous example the Dyson--Schwinger equation has the matrix form \eqref{two-f-DS} and again it is
reduced (cp. \eqref{label1}) to equation
\begin{equation}
\Big(\sum_{M=1}^{4}\mathcal{P}_{M}\bar{G}^{M}\Big)
\Big(\sum_{L=1}^{4}\mathcal{P}_{L}\bar{S}^{L}\Big)=
\mathcal{P}_{1}+\mathcal{P}_{2},
\end{equation}
where $\bar{G}_{M}$, $\bar{S}_{L}$ are now $2$-dimensional matrixes.

This yields the matrix analogies of equations \eqref{tf-DS-split}.
\begin{equation}
\begin{split}
\begin{array}{rcl}
\bar{G}_{1}\bar{S}_{1}+\bar{G}_{3}\bar{S}_{4} &=& E_{2},\\
\bar{G}_{2}\bar{S}_{2}+\bar{G}_{4}\bar{S}_{3} &=& E_{2},
\end{array}
\qquad\qquad
\begin{array}{rcl}
\bar{G}_{1}\bar{S}_{3}+\bar{G}_{3}\bar{S}_{2} &=& 0,\\
\bar{G}_{4}\bar{S}_{1}+\bar{G}_{2}\bar{S}_{4} &=& 0,
\end{array}
\end{split}
\end{equation}
where $E_{2}$ is a unit matrix $2\times 2$. It is easy to write down the solution of system
(that is the matrix analogy of \eqref{tf-DS-split-sol})
\begin{equation}
\label{p4-DS-split-sol}
\begin{split}
\begin{array}{rcl}
\bar{G}_{1} &=& \Big[\bar{S}_{1}-\bar{S}_{3}\big(\bar{S}_{2}\big)^{-1}\bar{S}_{4}\Big]^{-1},\\
\bar{G}_{2} &=& \Big[\bar{S}_{2}-\bar{S}_{4}\big(\bar{S}_{1}\big)^{-1}\bar{S}_{3}\Big]^{-1},
\end{array}
\qquad\qquad
\begin{array}{rcl}
\bar{G}_{3} &=& -\Big[\bar{S}_{1}-\bar{S}_{3}\big(\bar{S}_{2}\big)^{-1}\bar{S}_{4}\Big]^{-1}\bar{S}_{3}\big(\bar{S_{2}}\big)^{-1},\\
\bar{G}_{4} &=& -\Big[\bar{S}_{2}-\bar{S}_{4}\big(\bar{S}_{1}\big)^{-1}\bar{S}_{3}\Big]^{-1}\bar{S}_{4}\big(\bar{S_{1}}\big)^{-1}.
\end{array}
\end{split}
\end{equation}

Now let us concretize these general formulae. Suppose that we have
two fermions of different parity but there is no parity violation
in lagrangian. It means that the diagonal loops contain only the
$I$ and $\hat{p}$ matrixes.

\medskip
\hspace*{10mm}\includegraphics*{\imagepath{loop1}}\qquad \raisebox{\rbcorhh}{$\Sigma_{ii}\sim I,\;\hat{p}$,}
\medskip

while the non-diagonal ones have $\gamma^{5}$

\medskip
\hspace*{10mm}\includegraphics*{\imagepath{loop2}}\qquad \raisebox{\rbcorhh}{$\Sigma_{ij}\sim \gamma^5,\;\hat{p}\gamma^5$ %
for $i\neq j$}
\medskip

Therefore the decomposition of the inverse propagator in this basis looks as follows
\begin{equation*}
\begin{split}
S(p)=&\mathcal{P}_1  \left(
\begin{array}{cc}
-m_1+E-\Sigma^{(1)}_{11} & 0\\
 0 & -m_2+E-\Sigma^{(1)}_{22}
\end{array}\right)
+\\[3mm]
&+ \mathcal{P}_2 \left(
\begin{array}{cc}
-m_1-E-\Sigma^{(2)}_{11} & 0\\
0 & -m_2-E-\Sigma^{(2)}_{22}
\end{array}\right)
+\\[3mm]
&+ \mathcal{P}_3 \left(
\begin{array}{cc}
0 & -\Sigma^{(3)}_{12}\\
-\Sigma^{(3)}_{21} & 0
\end{array}\right)
+ \mathcal{P}_4 \left(
\begin{array}{cc}
0 & -\Sigma^{(4)}_{12}\\
-\Sigma^{(4)}_{21} & 0
\end{array}\right),
\end{split}
\end{equation*}
where $E=\sqrt{p^2}$, and $i,j=1,2$ numerate the dressing fermion states.

Substituting all into solution \eqref{p4-DS-split-sol},
we obtain the matrix of dressed propagator
\begin{equation}
\label{p4-DS-fv}
\begin{split}
G=&\Lambda^{+}\left(
\begin{array}{cc}
\dfrac{-m_2-E-\Sigma^{2}_{22}}{\Delta_1} & 0\\
 0 & \dfrac{-m_1-E-\Sigma^{2}_{11}}{\Delta_2}
\end{array}\right)+
\Lambda^{-}\left(
\begin{array}{cc}
\dfrac{-m_2+E-\Sigma^{1}_{22}}{\Delta_2} & 0\\
 0 & \dfrac{-m_1+E-\Sigma^{1}_{11}}{\Delta_1}
\end{array}\right)+                                 \\[3mm]
+&\Lambda^{+}\gamma^5\left(
\begin{array}{cc}
 0 & -\dfrac{\Sigma^{3}_{12}}{\Delta_1}\\
-\dfrac{\Sigma^{3}_{21}}{\Delta_2} & 0
\end{array}\right)+
\Lambda^{-}\gamma^5\left(
\begin{array}{cc}
 0 & -\dfrac{\Sigma^{4}_{12}}{\Delta_2}\\
-\dfrac{\Sigma^{4}_{21}}{\Delta_1} & 0
\end{array}\right).
\end{split}
\end{equation}
Here
\begin{equation*}
\begin{split}
\Delta_{1}&=\big(-m_1+E-\Sigma^{2}_{11}\big)\big(-m_2-E-\Sigma^{2}_{22}\big)-\Sigma^{3}_{12}\Sigma^{4}_{21},\\
\Delta_{2}&=\big(-m_1-E-\Sigma^{1}_{11}\big)\big(-m_2+E-\Sigma^{1}_{22}\big)-\Sigma^{4}_{12}\Sigma^{3}_{21}=
\Delta_{1}\big(E\to-E\big).
\end{split}
\end{equation*}

Let us compare \eqref{p4-DS-fv} with the dressing of two fermions of the same parity.
For this purpose we will rewrite the above solution \eqref{tf-one-par} in similar form
\begin{equation}
\label{p2-DS-fv}
\begin{split}
G&=\Lambda^{+}\frac{1}{\Delta_{1}}\left(
\begin{array}{cc}
-m_2+E-\Sigma^{1}_{22} & -\Sigma^{1}_{12}\\
-\Sigma^{1}_{21} & -m_1+E-\Sigma^{1}_{11}
\end{array}\right)
+\Lambda^{-}\frac{1}{\Delta_2}\left(
\begin{array}{cc}
-m_2-E-\Sigma^{2}_{22} & -\Sigma^{2}_{12}\\
-\Sigma^{2}_{21} & -m_1-E-\Sigma^{2}_{11}
\end{array}\right)               \\[3mm]
\Delta_1&=\big(-m_1+E-\Sigma^{1}_{11}\big)\big(-m_2+E-\Sigma^{1}_{22}\big)-\Sigma^{1}_{12}\Sigma^{1}_{21}\\
\Delta_2&=\big(-m_1-E-\Sigma^{2}_{11}\big)\big(-m_2-E-\Sigma^{2}_{22}\big)-\Sigma^{2}_{12}\Sigma^{2}_{21}=
\Delta_1\big(E\to-E\big).
\end{split}
\end{equation}

Let us remind that the both cases \eqref{p4-DS-fv}, \eqref{p2-DS-fv} correspond to conservation of
parity in lagrangian. The appearance  of nilpotent operators in decomposition \eqref{p4-DS-fv} is
an indication for transitions between states of different parity. They are absent in case of mixing
of the same parity states. Besides the denominators have different structure.

Let us summarize our consideration of the dressing of Dirac fermions.
\begin{enumerate}
\renewcommand{\labelenumi}{\theenumi)}
\item We found very convenient the using of the projection operators
      $\Lambda^{\pm}=\big(\sqrt{p^2}\pm\hat{p}\big)\big/2\sqrt{p^2}$ for solving of Dyson-Schwinger equation especially
      in case of few dressing states.
\item The projection operators $\Lambda^{\pm}$ are very useful in another aspect: its coefficients have the definite
      parity. But as one can see from the loop calculations \eqref{im-ofloop}, \eqref{rec-of-loop} the components
      $\Lambda^{\pm}$ have different parity. There is such correspondence: the parity of the field $\Psi$ is the parity
      of the component $\Lambda^{+}$, which has the pole $1\big/(E-m)$. Another component $\Lambda^{-}$ which has
      the pole $1\big/(-E-m)$ demonstrates the opposite parity.
\item In contrast to boson case, even if the interactions conserve the parity, the loop transitions between different
      parity states are not zero: they are proportional to nilpotent operator $\mathcal{P} \mathcal{P}=0$.
\item The joint dressing of two fermions without parity violation in vertex has different picture in dependence of
      parities of dressing states. One can illustrate it in the following scheme:
      \begin{equation*}
      \begin{array}{ccc}
      \mathparity{J^{P}}=1/2^{\pm} &\Leftrightarrow&
      \mathparity{J^{P}}=1/2^{\pm}\\
      &&\\
      \Lambda^{+}&\Longleftrightarrow&\Lambda^{+}\\
      \Lambda^{-}&\Longleftrightarrow&\Lambda^{-}
      \end{array}
      \qquad\qquad
      \begin{array}{ccc}
      \mathparity{J^{P}}=1/2^{\pm} &\Leftrightarrow&
      \mathparity{J^{P}}=1/2^{\mp}\\
      &&\\
      \Lambda^{+}&\Longleftrightarrow&\Lambda^{-}\\
      \Lambda^{-}&\Longleftrightarrow&\Lambda^{+}
      \end{array}
      \end{equation*}
      Another difference is the appearance of nilpotent operators in the second case.
\end{enumerate}

\section{Spin-parity of the Rarita--Schwinger field components}

Comparing the Tables~\ref{tt} and \ref{mult1} one can conclude that the presence of nilpotent operators
$\mathcal{P}_7$---$\mathcal{P}_{10}$ in the decomposition \eqref{ddecomp} of the Rarita--Schwinger propagator is an
indication for transitions between components of different parity $1/2^{\pm}$. To make sure in this conclusion we can
calculate the loop contributions in propagator. As an example we take the standard $\pi N\Delta$ interaction lagrangian
\begin{equation}
L_{int}=g_{\pi N\Delta}\bar{\Psi}\vphantom{\Psi}^{\mu}(x)\big(g^{\mu\nu}+a\gamma^{\mu}\gamma^{\nu}\big)
\Psi(x)\cdot\partial_{\nu}\phi(x) + h.c.\ .
\end{equation}
where the vertex contains an additional  parameter $a$.%
\footnote{%
To maintain the symmetry of free lagrangian \eqref{symm} parameter $a$ in vertex must be related with parameter $A$
in the free lagrangian $a=A(1+4z)/2+z$ (see, \cite{NEK}). Here $z$ is so called "off-shell" parameter the
meaning of which is controversial.
}

The one-loop self-energy contribution is
\begin{center}
\includegraphics*[height=2cm]{\imagepath{loop-rs}}
\end{center}
\vspace*{-5mm}
\begin{equation}
J^{\mu\nu}(p)=-ig_{\pi N\Delta}^2 \int \frac{d^4 k}{(2\pi)^4}
(g^{\mu\rho}+a\gamma^{\mu}\gamma^{\rho})k^{\rho}\frac{1}{\hat{p}+\hat{k}-m_N}
k^{\lambda} (g^{\lambda\nu}+a\gamma^{\lambda}\gamma^{\nu})
\frac{1}{k^2-\mpi^2}.
\end{equation}
Let us calculate the discontinuity of loop contribution in $\hat{p}$ basis \eqref{p-expan}.
\begin{equation}
\label{disc}
\begin{split}
\Delta J_1&=-ig^2 I_0\frac{m_N}{12s}\lambda(s,m_N^2,\mpi^2),  \\
\Delta J_2&=-ig^2 I_0\frac{1}{24s^2}(s+m_N^2-\mpi^2)\lambda , \\
\Delta J_3&=-ig^2 I_0\frac{m_N}{12s}(\lambda +6a\lambda -36a^2\mpi^2s),  \\
\Delta J_4&=-ig^2 I_0\frac{1}{24s^2}[(s+m_N^2-\mpi^2)\lambda +12as\lambda + 36a^2s(s^2-\mpi^2s-2m_N^2s-\mpi^2m_N^2+m_N^4)], \\
\Delta J_5&=ig^2 I_0\frac{m_N}{4s}[(s-m_N^2+\mpi^2)^2+2a(s-m_N^2+\mpi^2)^2+4a^2\mpi^2s],  \\
\Delta J_6&=ig^2 I_0\frac{1}{8s^2}[(s+m_N^2-\mpi^2)(s-m_N^2+\mpi^2)^2+
4as(s-m_N^2+\mpi^2)(s-m_N^2-\mpi^2)+         \\
&+ 4a^2s(s^2-\mpi^2s-2m_N^2s-\mpi^2m_N^2+m_N^4)],\\
\Delta J_7&=ig^2 I_0\sqrt{\frac{3}{s}}\cdot\frac{1}{24s}
[(s-m_N^2+\mpi^2)\lambda +4as(2s^2-\mpi^2s-4m_N^2s+2m_N^4-m_N^2\mpi^2 - \mpi^4)+ \\
&+ 12a^2s(s^2-\mpi^2s-2m_N^2s-m_N^2\mpi^2+m_N^4)], \\
\Delta J_8&=-ig^2 I_0\sqrt{\frac{3}{s}}\cdot\frac{a m_N}{6s}
[(s^2+4\mpi^2s-2m_N^2s+m_{N}^4-2m_{N}^2\mpi^2+\mpi^4)+6as\mpi^2], \\
\Delta J_9&=\Delta J_7                             \\
\Delta J_{10}&=-\Delta J_8.
\end{split}
\end{equation}
Here $I_0$ is the base integral \eqref{ibase}, $\lambda(a,b,c)=(a-b-c)^2-4bc$, arguments of $\lambda$ are the same
in all expressions, but are indicated only in first one.

We observed in case of Dirac fermions that the  propagator
decomposition in basis of projection operators demonstrates the
definite parity. We can expect the similar property for
Rarita--Schwinger field in $\Lambda$-basis. Let us verify it by
calculating the threshold behavior of imaginary part. Using
\eqref{disc}, one can convince yourself that
\begin{align*}
\Delta \bar{J}_1&= \Delta J_1 + E \Delta J_2 \sim q^3
\quad &
\Delta \bar{J}_3&= \Delta J_3 + E \Delta J_4 \sim q^3
\quad &
\Delta \bar{J}_5&= \Delta J_5 + E \Delta J_6 \sim q
\\
\Delta \bar{J}_2&= \Delta J_1 - E \Delta J_2 \sim q^5
\quad &
\Delta \bar{J}_4&= \Delta J_3 - E \Delta J_4 \sim q
\quad &
\Delta \bar{J}_6&= \Delta J_5 - E \Delta J_6 \sim q^3.
\end{align*}
Such a behavior indicates that the components $\bar{J}_1,\bar{J}_2$
exhibit the spin-parity%
\footnote{%
More exactly $\bar{J}_1 \sim 3/2^{+}$, $\bar{J}_2 \sim
3/2^{-}$, we found in above that the field parity is in fact the parity of
$\Lambda^{+}$ component in propagator.
}
$3/2^+$, while the pairs of coefficient $\bar{J}_3,\bar{J}_4$ and $\bar{J}_5,\bar{J}_6$ correspond to
$1/2^+$, $1/2^-$ contributions respectively.

\section{Conclusion}

We obtained in closed analytic form the dressed propagator of
Rarita--Schwinger field taking into account all spin components.
To derive it we introduced the spin-tensor basis \eqref{L-basis}
with very simple multiplicative properties. It is constructed by
combining the standard tensor operators \eqref{present} and
$\Lambda^{\pm}$ projection operators \eqref{project}. This basis
consists of $6$ projection operators and $4$ nilpotent ones.

In search of analogy for dressing of $s=1/2$ sector of the
Rarita--Schwinger field we have considered the dressing of Dirac
fermions in different variants. As a result we have found that
the nearest analogy is the joint dressing of two Dirac fermions of
different parity. Using of projection operators $\Lambda^{\pm}$ is
turned out to be very convenient in this consideration. Besides
the simple multiplicative properties, this basis is convenient
also in another aspect: its coefficients exhibit the definite
parity. The presence of nilpotent operators in propagator
decomposition yields indication for transitions between different
parity states. Calculation of the self-energy contributions in
case of $\Delta$ isobar confirms it: in the Rarita--Schwinger
field besides the leading $s=3/2$ contribution there are also two
$s=1/2$ components of opposite parity. Similar conclusion was
obtained in \cite{Kir} on the base of algebraic methods and
construction of the corresponding space of the Lorentz group
representations.

The obtained dressed propagator \eqref{solve} solves an algebraic
part of the problem, the following step is renormalization. Note that
the investigation of the dressed propagator is an alternative for more
conventional methods based on equations of motion (see, \textit{i.e.}
Ref.~\cite{Pas98,Pas99} and references therein).
Instead of analysis of motion equations and constrains it leads to investigation of poles in complex energy plane.
However the renormalization problem of the Rarita--Schwinger field needs a more careful consideration.

\section*{Acknowledgments}

We thank to G. Lopez Castro and D. V. Ahluwalia-Khalilova for
references and comments and also to D. V. Naumov for useful
remarks. We are thankful to participants of JINR--ISU Baikal
summer school for fruitful discussions.

\appendix

\section*{\appendixname}

\section*{Renormalization of Dirac fermion propagator in $\Lambda^{\pm}$ basis}

The dressed inverse propagator has the form
\begin{equation}
S(p)=\hat{p}-m-A\big(p^2\big)-\hat{p}B\big(p^2\big)=-m-A\big(p^2\big)+\hat{p}\Big(1-B\big(p^2\big)\Big),
\label{a1}
\end{equation}
where $A(p^2)$ and $B(p^2)$ are components of self-energy $\Sigma(p)$
\begin{equation*}
\Sigma(p)=A(p^2)+\hat{p}B(p^2),
\end{equation*}

The standard procedure of renormalization makes use of decomposition of the inverse propagator in terms of $\hat{p}-m$.
If we use the mass shell subtraction scheme then we have the following condition
\begin{equation}
S(p)=\hat{p}-m+o(\hat{p}-m)\qquad\text{when}\quad(\hat{p}-m)\to 0
\label{a2}
\end{equation}
where $m$ is a renormalized fermion mass. It leads to conditions for coefficients $A$ and $B$.
\begin{equation}
\label{a3}
\begin{split}
A(m^2)+mB(m^2)=0, \\
2mA^{\prime}(m^2)+B(m^2)+2m^2B^{\prime}(m^2)=0
\end{split}
\end{equation}
To derive this conditions it is necessary to use the relation $p^2=\hat{p}^2$. The conditions \eqref{a3} define
the subtraction constants in the loops.

With using of $\Lambda^{\pm}$ projection operators basis the procedure
of the renormalization becomes slightly different. The inverse propagator has form
\begin{equation*}
S(p)=\Lambda^{+}S^{1}+\Lambda^{-}S^{2},
\end{equation*}
where
\begin{gather*}
S^{1}=-m-A(p^2)+\sqrt{p^2}\Big(1-B\big(p^2\big)\Big),\\
S^{2}=-m-A(p^2)-\sqrt{p^2}\Big(1-B\big(p^2\big)\Big)
\end{gather*}
One should renormalize the scalar functions $S^{1,2}$ depending on argument $E=\sqrt{p^2}$. Let us require
$S^{1}$ to have zero at the point $E=m$ with unit slope.
\begin{equation*}
S^{1}=E-m+o(E-m) \qquad\text{when}\quad E \to m
\end{equation*}

It is easy to see that this condition coincides with \eqref{a2} after substitution $\hat{p}\to\sqrt{p^2}$ and therefore we
get the same conditions \eqref{a3} on the subtraction constants.

This condition defines completely the subtraction constants of $A(p^2), B(p^2)$ so the another component
$S^{2}$ is fixed also. As a result we obtain the dressed renormalized propagator $G(p)$, which coincides with the
standard expression.

\end{document}